\documentclass[review]{elsarticle}
\usepackage{graphicx}
\usepackage{amsmath}
\usepackage{amssymb}
\usepackage{bm}
\usepackage{subfigure}
\usepackage{color}

\title{Multi-GPU-based Swendsen-Wang multi-cluster algorithm for the simulation 
of two-dimensional $q$-state Potts model} 
\author[tmu]{Yukihiro Komura}
\ead{y-komura@phys.se.tmu.ac.jp}
\author[tmu]{Yutaka Okabe}
\ead{okabe@phys.se.tmu.ac.jp}
\address[tmu]{Department of Physics, Tokyo Metropolitan University, Hachioji, Tokyo 192-0397, Japan}

\begin{document}

\begin{abstract}
We present the multiple GPU computing 
with the common unified device architecture 
(CUDA) for the Swendsen-Wang multi-cluster algorithm 
of two-dimensional (2D) $q$-state Potts model. 
Extending our algorithm for single GPU computing [Comp. Phys. Comm. 183 
(2012) 1155], we realize the GPU computation of 
the Swendsen-Wang multi-cluster algorithm for multiple GPUs.  
We implement our code on the large-scale open science supercomputer 
TSUBAME 2.0, and test the performance and the scalability of 
the simulation of the 2D Potts model. 
The performance on Tesla M2050 using 256 GPUs is obtained 
as 37.3 spin flips per a nano second for the $q=2$ Potts model 
(Ising model) at the critical temperature with 
the linear system size $L=65536$. 
\end{abstract}


\maketitle

\section{Introduction}
High performance computing accelerates advances in science.  
Recently the use of general purpose computing on graphics processing 
unit (GPU) is a hot topic in computer science. 
Drastic reduction of processing times can be realized in 
scientific computations. 
Using the common unified device architecture (CUDA) released by NVIDIA, 
it is now easy to implement algorithms on GPU 
using standard C or C++ language with CUDA specific extension. 

Multiple GPU computing is performed for accommodating 
larger-scale problems beyond the capacity of video memory 
on a single GPU and for speedup of the fixed problem 
pursuing strong scalability.  Multiple GPU computing 
requires GPU-level parallelization. 
A large-scale open science supercomputer TSUBAME 2.0 
is available at the Tokyo Institute of Technology. 
TSUBAME 2.0 consists of 4224 NVIDIA Tesla M2050 GPUs as a total, 
and the theoretical peak performance reaches 2.4 PFLOPS 

Studying spin models is a basic problem in statistical physics. 
At the same time the Monte Carlo simulation of spin models 
has been served as a benchmark of newly developed technology. 
Sublattice decomposition was used in the Metropolis-type \cite{metro53}
Monte Carlo simulation of spin models in vector computer, 
and the same idea was employed in the GPU-based calculation 
of the Ising model by Preis {\it et al.} \cite{preis09, preis11}.
The GPU-based calculation of the multispin 
coding of the Metropolis-type Monte Carlo simulation 
was proposed by Block {\it et al.} \cite{block10}.
There, they argued the multiple GPU calculation 
in order to overcome the memory limitations of a single GPU, and 
they found that for large systems, the computation time scales nearly linearly with 
the number of GPUs. 

The Metropolis-type single-spin-flip algorithm often suffers 
from the problem of slow dynamics or the critical slowing down. 
To overcome difficulty, a cluster flip algorithm was proposed 
by Swendsen and Wang \cite{sw87}.  In the Swendsen-Wang (SW) algorithm, 
all the spins are partitioned into clusters; 
thus, the SW algorithm is called the multi-cluster algorithm. 
Wolff \cite{wolff89} proposed another type of cluster algorithm, 
that is, a single-cluster algorithm, where only a single cluster 
is generated, and the spins of that cluster are updated. 

The high performance computing using GPUs are highly desirable 
for Monte Carlo simulations with cluster flip algorithms. 
Recently, some attempts have been reported along this line 
on a single GPU. 
The present authors \cite{komura11} have proposed the GPU-based 
calculation with CUDA for the Wolff single-cluster algorithm, 
where parallel computations are performed for the newly added spins 
in the growing cluster. 
Hawick {\it et al.} \cite{Hawick_single_cluster} have studied 
the CUDA implementation of the Wolff algorithm 
using a modified connected component labeling 
for the assignment of the cluster.  They put more emphasis on 
the hybrid implementation of Metropolis and Wolff updates and 
the optimal choice of the ratio of both updates. 
Weigel \cite{weigel11} has studied parallelization of 
cluster labeling and cluster update algorithms for calculations 
with CUDA. 
He realized the SW multi-cluster algorithm 
by using the combination of self-labeling algorithm and 
label relaxation algorithm or hierarchical sewing algorithm. 
Quite recently, the present authors \cite{komura12} have proposed the GPU 
calculation with CUDA for the SW multi-cluster algorithm 
by using the two connected component 
labeling algorithms, the algorithm by Hawick {\it et al.} 
\cite{Hawick_labeling} and that by Kalentev {\it et al.} 
\cite{Kalentev}, for the assignment of clusters. 
The computational speed for the $q=2$ Potts model
on NVIDIA GeForce GTX580 was 12.4 times as fast as
the computational speed on a current CPU core,
Intel(R) Xeon(R) CPU W3680. 

Finite-size scaling works well for the systems which undergo 
second-order phase transition; therefore numerical studies on 
smaller systems can lead to the understanding of phase transition 
with the help of finite-size scaling.  On the other hand, 
it is a different story for the Kosterlitz-Thouless (KT) transition 
\cite{KT}, where logarithmic behavior is essential.  
Large scale simulations are needed for the systems which show 
the KT transition. 

In this paper, we present the multiple GPU calculation with CUDA 
for the SW multi-cluster algorithm of the two-dimensional (2D) 
$q$-state Potts model.  This is an extension of the GPU-based calculation 
on a single GPU presented in Ref. \cite{komura12}.  
We test the performance of our algorithm for the 2D $q$-state Potts model 
on the multiple GPUs of TSUBAME 2.0.  
The rest of the paper is organized as follows. 
In section 2, we briefly review the GPU-based calculation of the SW 
multi-cluster algorithm on a single GPU described in Ref. \cite{komura12}. 
In section 3, we present the idea and implementation of GPU-based 
calculation extended to the case of multiple GPUs. 
In section 4, we test the performance of multiple GPU calculation 
on TSUBAME 2.0. 
The summary and discussion are given in section 5. 

\section{Single GPU computing of the Swendsen-Wang cluster algorithm}

Swendsen and Wang proposed a Monte Carlo algorithm 
of multi-cluster flip \cite{sw87}. 
To explain the SW algorithm, we use the $q$-state Potts model 
whose Hamiltonian is given by
\begin{equation}
 \mathcal{H} = -J \sum_{<i,j>}(\delta_{S_{i},S_{j}}-1), 
              \quad S_{i} = 1, 2, \cdots, q, 
\end{equation}
and this corresponds to the Ising model for $q$ = 2.
Here, $J$ is the coupling and $S_{i}$ is the Potts spin 
on the lattice site $i$. The summation is taken over 
the nearest neighbor pairs $<i,j>$.  
Periodic boundary conditions are employed. 

There are three main steps in the SW algorithm:
(1) Construct a bond lattice of active or non-active bonds 
depending upon the temperature. 
(2) The active bonds partition the spins into clusters which are 
identified and labeled using a cluster-labeling algorithm. 
(3) All spins in each cluster are set randomly to one of $q$.

For an efficient cluster-labeling algorithm, 
the Hoshen-Kopelman algorithm \cite{Hoshen_Kopelman}, which was 
first introduced in context of cluster percolation, is 
often used.
In the Hoshen-Kopelman cluster-labeling algorithm, integer labels 
are assigned to each spin in a cluster. Each cluster has 
its own distinct set of labels.  The proper label of a cluster, 
which is defined to be the smallest label of any spin in the cluster, 
is found by the following function. 
The array \verb+label+ is used, and 
if \verb+label+ is a label belonging to a cluster, 
\verb+label[label]+ is the index of another label 
in the same cluster which has a smaller value 
if such a smaller value exists.  The proper label for the cluster 
is found by evaluating \verb+label[label]+ repeatedly. 

Since the calculations of the step of active bond generation and
the step of spin flip are done independently on each site, these
steps are well suited for parallel computation on GPU. On the other
hand, in the step of cluster labeling the assignment of label of
cluster is done on each site piece by piece sequentially; thus the
cluster-labeling algorithm such as the Hoshen-Kopelman algorithm
cannot be directly applied to the parallel computation on GPU.

The present authors \cite{komura12} have proposed the calculation 
on a single GPU with CUDA for the SW multi-cluster algorithm 
by using the cluster-labeling algorithm appropriate for 
the parallel computation using GPU.  The recently proposed 
algorithm, the "Label Equivalence" algorithm, by Hawick {\it et al.} 
\cite{Hawick_labeling} together with its refined version 
by Kalentev {\it et al.} \cite{Kalentev} were used 
in Ref. \cite{komura12}.  
Since the algorithm by Kalentev {\it et al.} \cite{Kalentev} is superior to 
that by Hawick {\it et al.} \cite{Hawick_labeling}, and we extend it 
to the multiple GPU computing, we briefly explain the "Label Equivalence" 
algorithm by Kalentev {\it et al.} \cite{Kalentev}. 

The procedure of their algorithm \cite{Kalentev}, which was 
schematically illustrated in Fig. 2 of Ref. \cite{komura12}, consists of 
the two kernel functions, that is, scanning function and analysis function. 
The scanning function compares the label of each site with 
that of the nearest-neighbor sites when the bond between 
each site and the nearest-neighbor site is active. 
If the label of the nearest-neighbor site is smaller than the label 
of that site, the label variable with the label number, 
\verb+label[label]+, is updated to the smallest one. 
The analysis function resolves the equivalence chain of the label obtained 
in the scanning function; the label variable 
is updated from the starting site 
to the new site, which is similar to the method of 
the Hoshen-Kopelman algorithm. 
Each thread tracks back the label variable until 
the label variable, \verb+label+, remains unchanged. 
In the "Label Equivalence" algorithm, 
the loop including two kernel functions is iterated up to the point 
when the information of the labeling needs 
no more process of scanning function. 
The number of iterations for the $q=2$ Potts model 
at the critical temperature 
is about 8.6 on average for 4096$\times$4096 systems.

\section{Multiple GPU computing of the Swendsen-Wang cluster algorithm}

To use multiple GPUs, the data communication between GPUs is necessary. 
However it is not able to communicate directly between GPUs 
when many GPUs are used. 
The data transfers on three levels are needed, that is, the data transfer 
from GPU to CPU, that between CPUs and that from CPU to GPU. 
We use the message passing interface (MPI) library for the communication 
between CPUs, and we run the same process number for MPI as the GPU number. 
These data communications reduce the computation efficiency of multiple GPUs. 
Thus, hiding the cost of data communication by overlapping communication 
and computation is desirable such as in 
Ref. \cite{Aoki_TSUBAME,Shimokawabe_TSUBAME2010,Shimokawabe_TSUBAME}.

\begin{figure}
\begin{center}
\includegraphics[width=0.8\linewidth]{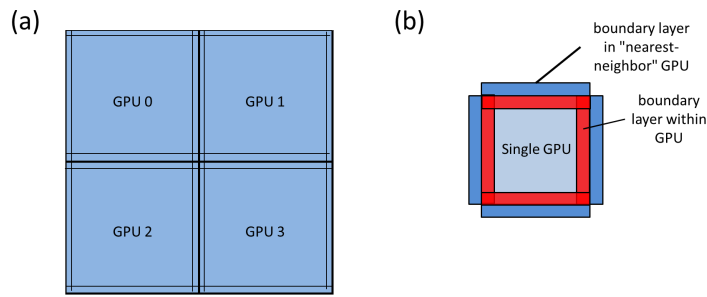}
\caption{
(a) The $2 \times 2$ super-lattice structure for 4 GPUs.  
(b) The information on a single GPU.  It has the arrays to preserve 
the data of surrounding boundary layers 
and to transfer the data of boundary layers.
}
\label{fig:fig1}
\end{center}
\end{figure}

We arrange the total lattice with a super-lattice structure 
using multiple GPUs.  Each GPU has the information of spins 
on a sub-lattice together with the arrays to preserve the data of 
surrounding boundary layers and to transfer the data of boundary layers.  
We illustrate the case of $2 \times 2$ super-lattice structure using 4 GPUs 
in figure \ref{fig:fig1}. 
The present super-lattice and sub-lattice structures are similar to the 
structures employed in \cite{block10}. 

The calculation of active bond generation is performed in the same manner 
as that for a single GPU only with the copy of boundary layers. 
However, the calculation of cluster labeling needs extra process. 
Within each GPU, we follow the cluster labeling method of 
Kalentev \cite{Kalentev}, which consists of the scanning kernel function 
and analysis kernel function.  But we have to check whether the sites of 
the different sub-lattices (GPUs) belong to the same cluster.  
We employ a two-stage process of cluster labeling.  
After the cluster labeling within each GPU is finished, 
we check the bond between the sites of "nearest-neighbor" GPU; 
that is, the inter-GPU scanning function.  
For this purpose, we use two variables to describe the labeling 
of the cluster.  
One is a temporal variable \verb+label+ to make a cluster labeling 
within each GPU.  The other variable \verb+label_total_lattice+ 
has the information on the labeling over the total lattice; 
it is described by the combination of the GPU number and 
the labeling within that GPU.  When the labeling of 
the "nearest-neighbor" GPUs is compared, 
the variable \verb+label_total_lattice+[\verb+label+] 
is updated to the  \verb+label_total_lattice+ 
with smaller GPU number. 
Then, the equivalence chain is resolved in a similar method to 
the analysis kernel function within a single GPU; that is, 
the inter-GPU analysis function. 

To accelerate the computational speed, we 
arrange the process 
of the inter-GPU cluster labeling such that only the calculation 
of boundary layers in the sub-lattice is made. 
In the inter-GPU scanning function, we make parallel calculation 
for the boundary layers in "nearest neighbor" GPU. 
In the inter-GPU analysis function, we only update 
the variable 
\verb+label_total_lattice+[\verb+label+] 
of the arrays to transfer the data of boundary layers within GPU. 
After the iteration of the inter-GPU cluster labeling ends, 
each GPU updates the variable 
\verb+label_total_lattice+[\verb+label+] 
of the total sub-lattice. 
By using this method, 
the amount of calculation in the inter-GPU cluster labeling 
is proportional to $4N_x$ instead of $N_x \times N_x$ 
for the linear system size of each GPU $N_x$. 

The step of spin update is not straightforward. 
If we try to choose a new spin state after the step of cluster labeling, 
we face with the problem that we have to copy all the information of 
new spin states. 
To overcome this difficulty, we include the information on new spin state 
in the process of cluster labeling.  
Because of distributed memory, we cannot directly access the spin state 
of different GPU. 
This problem of multi-cluster flip Monte Carlo is not a local one.  
The information of boundary layer is not enough. 
The situation is different from the case of partial differential equation, 
for example. 

The informations of new spin state and active bonds 
with up and left directions are included in a single word 
\verb+label+ or \verb+label_total_lattice+ 
to save memory and access time. 
The data structure is shown in figure {\ref{fig:fig1b}. 
The \verb+label+ is defined as an unsigned integer type of 32 bit, and 
the \verb+label_total_lattice+ is defined as an unsigned 
long long integer type of 64 bit. 
We note that the new spin state in \verb+label_total_lattice+ 
is that for the labeling in the total lattice, and 
is not necessarily the same as the temporal new spin state in \verb+label+.

\begin{figure}
\begin{center}
\includegraphics[width=0.8\linewidth]{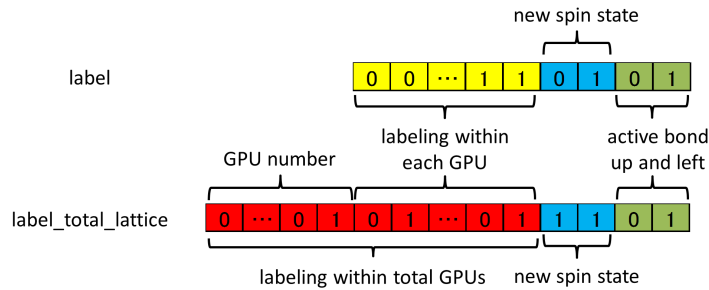}
\caption{
The data structures of two variables for labeling. 
A temporal variable "label" is used for saving the labeling within 
each GPU, and a variable "label\_total\_lattice" is used 
for saving the labeling information on the total lattice. 
The informations on new spin state and active bonds of up and 
left directions are included in a single word.  
The case of 4-state spin is shown as an example. 
}
\label{fig:fig1b}
\end{center}
\end{figure}

The data communication between GPUs is needed only 
in the step of cluster labeling. 
In order to improve the performance of the multiple-GPU computing, 
we use an overlapping technique between communication 
and computation.  We pick up the data which are not referenced 
in the running kernel function. 
The scanning kernel function does not reference the data 
for the transfer of boundary layers within each GPU, and 
the analysis kernel function does not reference the data 
of boundary layers in "nearest-neighbor" GPU. 
The scanning kernel function is launched as the stream 0; 
the data for the transfer 
of boundary layers within each GPU are transferred 
from the global memory to host memory using the asynchronous 
Application Programming Interface (API) as the stream 1. 
In a similar way, the analysis kernel function is launched 
as the stream 0, and the data of boundary layers 
in "nearest-neighbor" GPUs are 
transferred from the host memory to global memory 
using the asynchronous API as the stream 1. 
We use the technique of overlapping communication 
and computation for both the cluster labeling within each GPU 
and the inter-GPU cluster labeling. 
We note that it needs two loops of scanning function 
and analysis function to reflect the update of data 
of boundary layers in "nearest-neighbor" GPU.
To terminate the loop in the cluster labeling, 
we check that the termination conditions for all GPUs 
are satisfied.  To do this, the "flag"s for the termination condition 
are gathered in one CPU 
during the implementation of analysis kernel function. 
And that CPU checks the total termination condition 
and the check result of termination is communicated to all CPUs by MPI command. 
The procedure of overlapping data transfer 
and computation at the step of cluster labeling 
is schematically shown in figure \ref{fig:fig2}. 

We finally note that we use a linear congruential random generator  
which was proposed by Preis {\it et al.} \cite{preis09} 
when random numbers are generated.

\begin{figure}
\begin{center}
\includegraphics[width=0.8\linewidth]{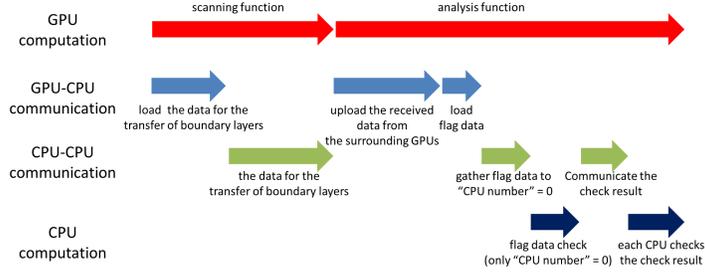}
\caption{
The procedure of overlapping data transfer 
and computation at the step of cluster labeling. 
During the computation in the scanning function, 
the data of boundary layers are 
transferred from each GPU to each CPU, and then 
the data are transferred from each CPU 
to the surrounding CPUs by MPI command. 
During the computation in the analysis function, 
the received data from the surrounding CPUs
are uploaded from each CPU to each GPU.  
To check the termination conditions in all GPUs, 
the "flag"s of termination condition for all GPUs 
are gathered in one CPU, and that CPU 
checks the total termination condition and 
communicates the result of termination to all CPUs. 
}
\label{fig:fig2}
\end{center}
\end{figure}

\section{Results}

We have implemented our code of the multiple GPU computing 
for the 2D $q$-state Potts models on TSUBAME 2.0 
with NVIDIA Tesla M2050 GPUs using CUDA 4.0 and openMPI 1.4.2. 
We have checked our code by plotting the temperature dependence 
of the moment ratio of magnetization as in Fig. 4 of 
Ref. \cite{komura12}, for example. 
We here present the data of the performance for the $q=2$ Potts model
 (Ising model) at the critical temperature, 
$T_{c}/J = 1/\ln(1+\sqrt{2}) = 1.1346$. 
The average amount of spin flips per a nano second 
at the critical temperature for the $q=2$ Potts model 
is tabulated in table \ref{tb:GPU_time_q=2_Potts}. 
There, 
the average amounts of spin flips per a nano second 
for only spin updates and 
that including the measurement of energy and magnetization are given.
The size of sub-lattice for each GPU is fixed as $4096 \times 4096$, 
and we increase the number of GPUs. 
The linear system sizes for the total lattice are 
$L$=4096, 8192, 16384, 24576, 32768, 49152 and 65536 
for the number of GPUs as 1, 4, 16, 36, 64, 144 and 256, respectively. 

\begin{table*}[htbp]
\begin{center}
\begin{tabular}{llll}
\hline
&  system size of total lattice $L$  & update only       & update  \\
&  \ (the number of GPUs )           &                   & \ + measurement \\
\hline
& 4096  (  1 GPU)        & 0.232 & 0.206 \\ 
& 8192  (  4 GPU)        & 0.813 & 0.729 \\
& 16384 ( 16 GPU)        & 2.927 & 2.657 \\
& 24576 ( 36 GPU)        & 6.149 & 5.582 \\
& 32768 ( 64 GPU)        & 10.36 & 9.599 \\
& 49152 (144 GPU)        & 22.67 & 20.53 \\
& 65536 (256 GPU)        & 37.30 & 33.85 \\
\hline
\end{tabular}
\caption{
Average amount of spin flips per a nano second 
at $T_c$ for the $q=2$ Potts model with multiple GPUs. 
The numbers of spin flips per a nano second 
for only spin updates and that including the measurement 
of energy and magnetization are given.}
\label{tb:GPU_time_q=2_Potts}
\end{center}
\end{table*}

\begin{figure}
\begin{center}
\includegraphics[width=0.8\linewidth]{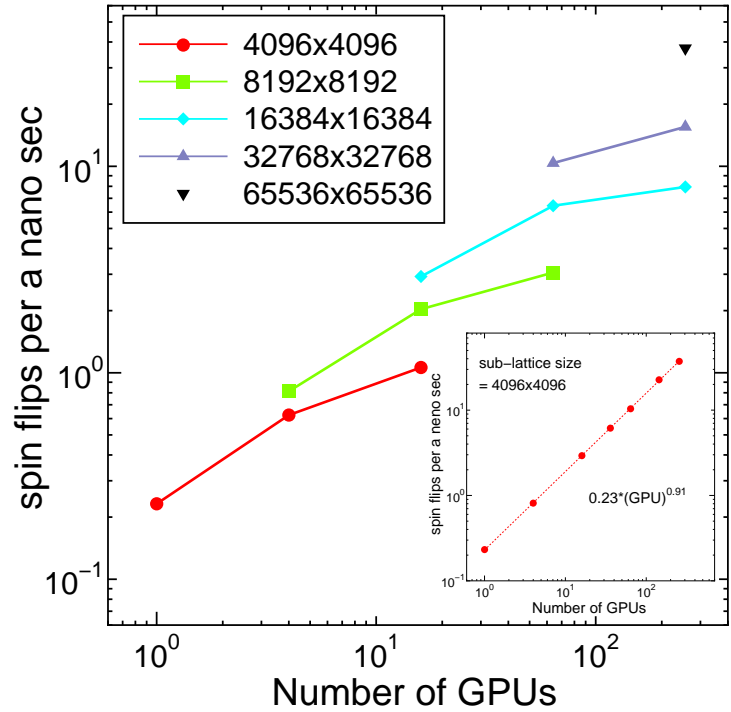}
\caption{
Plot of spin flips per a nano second with changing the number of GPUs  
at the critical temperature for the $q=2$ Potts model.  
The data for the sub-lattice size for each GPU as 
$1024 \times 1024$, $2048 \times 2048$ and $4096 \times 4096$ 
are plotted.  The same mark is used for plotting the data 
for fixing the linear system size of the total lattice as 
$L$=4096, 8192, 16384, 32768 and 65536. 
In the inset the values of fixing the sub-lattice size as $4096 \times 4096$ 
are plotted as a function of the number of GPUs. The best-fitted curve 
to get the power dependence is given by the dotted line.
}
\label{fig:fig3}
\end{center}
\end{figure}
To compare with the code for single GPU computing, we run the code 
in Ref. \cite{komura12} with NVIDIA Tesla M2050. 
The average amount of spin flips per a nano second at $T_c$ 
for the $q=2$ Potts model of size $4096 \times 4096$ is 
0.26. 
Thus, the calculation speed of the code for multi GPU computing 
in this paper, 0.23 spin flips per a nano second, 
is nearly equal to the calculation speed of the code for single GPU computing. 
And, the performance of our code for multi GPU computing 
increases with the number of GPUs. 
The best performance is 37.3 spin flips per a nano second for the system 
with $L=65536$ by using $256$ GPUs. 

Next we examine the scalability of our code 
with changing the number of GPUs 
at the critical temperature for the $q=2$ Potts model. 
We give the double logarithmic plot of the average amount of spin flips 
per a nano second as a function of 
the number of GPUs in figure \ref{fig:fig3}. 
For the size of sub-lattice of each GPU, not only the data 
for $4096 \times 4096$ but also those for $2048 \times 2048$ 
and $1024 \times 1024$ are plotted.  
The dependences on the number of GPUs with fixing 
the total linear system size as $L=4096, 8192, 16384, 32768, 65536$ 
are shown by using the same mark in figure \ref{fig:fig3}. 
Since the number of iterations at the step of inter-GPU 
cluster labeling increases with the increase of the number of GPUs,  
the efficiency of the average spin flips 
per a nano second slightly decreases with the decrease of the sub-lattice size. 
We can see this tendency in Figure \ref{fig:fig3}. 
In the inset the dependence on the number of GPUs 
with fixing the sub-lattice size as $4096 \times 4096$ is shown. 
We see that the performance of our code 
increases as a power of the number of GPUs. 
The coefficient of the power is estimated as 0.91 
by the best-fitted curve shown in the inset. 
Since it takes extra time to diffuse the label number 
from one GPU to all GPUs, 
the perfect weak scalability, that is, 
the power of 1.0, is not achieved, 
but we can say the weak scalability is well satisfied 
for our code.

\section{Summary and discussion}

Extending our algorithm for single GPU computing proposed 
in Ref. \cite{komura12}, 
we have developed the calculation of the SW multi-cluster algorithm 
for multiple GPU computing.  
In the step of cluster labeling, we employ a two-stage process; 
that is, we first make the cluster labeling within each GPU, 
and then the inter-GPU labeling is performed 
with reference to the label on the first-stage. 
We use some trick for the step of spin update; we include the information 
on new spin state in the process of cluster labeling. 
A special attention is paid to hide the communication overhead. 
We use the technique of overlapping communication and computation. 
And, we arrange the process 
of the inter-GPU cluster labeling such that only the calculation 
of boundary layers in the sub-lattice is made in order to 
gain the computational speed. 
The multiple GPU computing can be used not only for higher 
performance but also for accommodating larger-scale problems 
beyond the capacity of video memory on a single GPU. 
Thus the algorithm proposed in this paper permits us to deal with 
the calculation of the SW multi-cluster Monte Carlo simulation  
for large systems. 

We have implemented our code of the SW algorithm 
for the 2D $q$-state Potts model 
on the large-scale open science supercomputer TSUBAME 2.0. 
We have tested the performance of our code for the $q=2$ 
Potts model. 
As a result, the average amount of spin flips per a nano second 
by using 256 GPUs is 37.3 
at the critical temperature 
for the total linear system size $L=65536$. 
And we have examined the scalability of our code at $T_c$ 
with changing the number of GPUs.   
The average amount of spin flips per a nano second 
is proportional to a power of the number of GPUs 
with the power being 0.91. 
We have confirmed both the strong and weak scalabilities.

We have shown in this paper that the SW multi-cluster algorithm 
is well suited for multiple GPU computing, although an early study 
suggested that the SW multi-cluster algorithm is not suitable for 
the multiple parallel computation \cite{Barkema94}. 
We emphasize that the implementation of the cluster algorithm
for multiple GPU computing is highly efficient
because the convergence of the cluster algorithm is several orders
of magnitude faster than the single-spin flip Monte Carlo simulation
near the critical point for large systems. 
This computation is easily extended to continuous spin models, 
such as the classical XY model, using the Wolff's idea of 
embedded cluster \cite{wolff89}.  
It is interesting to perform large scale simulations 
for the 2D classical XY model, which undergoes the KT transition, 
as mentioned in Introduction. 
The multiple GPU calculation for the cluster algorithm of 
the classical XY model in connection with 
the probability-changing cluster algorithm \cite{tomita2002a,tomita2002b} 
is now in progress.

\section*{Acknowledgment}
We thank Takayuki Aoki for valuable discussions. 
The computation of this work has been done using 
the large-scale open science supercomputer TSUBAME 2.0 
at the Tokyo Institute of Technology. 
This work was supported by a Grant-in-Aid for Scientific Research from
the Japan Society for the Promotion of Science.

\end{document}